\documentclass[12pt]{article}

\usepackage{scicite}
\usepackage{graphicx}
\usepackage{url}
\usepackage{times}
\usepackage{color}

\newenvironment{sciabstract}{%
\begin{quote} \bf}
{\end{quote}}

\title{Spatial Aggregation Distorts COVID-19 Growth Rates}
\title{Unequal Impact and Spatial Aggregation Distort COVID-19 Growth Rates}

\author
{Keith Burghardt$^{1\ast}$ and Kristina Lerman$^{1}$ \\
\\
\normalsize{$^{1}$Information Sciences Institute, University of Southern California,}\\
\normalsize{4676 Admiralty Way, Marina del Rey, CA 90292, USA}\\
\normalsize{$^\ast$To whom correspondence should be addressed; E-mail:  keithab@isi.edu.}
}

\date{\today}

\begin{document} 


\baselineskip24pt


\maketitle


\begin{sciabstract}
The COVID-19 pandemic has emerged as a global public health crisis. To make decisions about mitigation strategies and to understand the disease dynamics, policy makers and epidemiologists must know how the disease is spreading in their communities. We analyze confirmed infections and deaths over multiple geographic scales to show that COVID-19's impact is highly unequal: many subregions have nearly zero infections, and others are hot spots. 
We attribute the effect to a Reed-Hughes-like mechanism in which disease arrives at different times and grows exponentially. 
Hot spots, however, appear to grow faster than neighboring subregions and dominate spatially aggregated statistics, thereby amplifying growth rates. 
The staggered spread of COVID-19 can also make aggregated growth rates appear higher even when subregions grow at the same rate. Public policy, economic analysis and epidemic modeling need to account for potential distortions introduced by spatial aggregation. 
\end{sciabstract}

\section*{Introduction}







The COVID-19 pandemic has spread rapidly around the globe, claiming hundreds of thousands of lives and wreaking havoc on world economies. Public health experts and policy makers must consider a complex array of metrics when deciding when and how to enforce mitigation strategies, such as closing schools and businesses. An important consideration in these calculations is a measure of how quickly the virus is spreading within the communities: 
a fast spreading virus may force municipalities, states and nations to order residents to shelter at home to slow transmission. 
Epidemiologists must similarly measure the growth rate to better understand the underlying mechanism of the disease, including its basic reproduction number \cite{Delamater2019}, or to forecast its spread~\cite{IHME2020}. 

Using confirmed infections and deaths data from a variety of sources around U.S. and the world, we show that the impact of COVID-19 is highly unequal, with \textit{hot spots} emerging at multiple spatial scales \cite{Lessler2017}: from individual facilities \cite{NYTclusters} and city neighborhoods \cite{LADPH}, to U.S. counties and states~\cite{NYT2020}, to nations~\cite{OWD2020}. We also show that spatial aggregation of COVID-19 data leads to higher growth rates than within most subregions, which we call \textit{aggregation bias}.
As a result,  growth rate at a city-level overestimates how quickly the disease spreads through city neighborhoods, and state-level growth rates are higher than for most counties within each state. 

We argue that hot spots and aggregation bias arise 
because disease appears in new subregions at different times and grows at different rates. As a result, subregions where the disease is spreading more quickly grow to become hot spots and dominate statistics. Spatial aggregation of data over the subregions produces growth estimates that are systematically higher  relative to growth rates within most subregions. More interestingly, because disease arrives in subregions at different times, spatial aggregation can also exaggerate the growth rate even when disease is spreading at the same rate within the subregions.

To better understand aggregation bias, we create a simple stochastic model that is variant of a Reed-Hughes mechanism~\cite{Reed2002}, with synthetic communities in which the disease arrives at different times and grows at different rates
. 
Both the arrival times and growth rates are picked from empirical distributions. We show that the varying ages of outbreaks create a heavy-tailed distribution of the number of infections and deaths, with a small number of hot spots representing the majority of all infections and deaths. The size of the outbreak is highly correlated with the growth rate in the subregion; therefore, when the synthetic data is aggregated to simulate state or national statistics, these hot spots systematically amplify the estimated growth rates, much like what is observed empirically. In addition, even when growth rates are the same, the staggered arrival of the virus in communities amplifies the growth rates in aggregated data.


Epidemic modeling and public health policy need to account for the role 
biases play in data aggregation. When calculating the costs and benefits of lock-downs, for example, analysts must control for these biases to better understand the risks people face. Aggregate data could also affect parameters in epidemic models and therefore reduce model prediction accuracy.


\section*{Results}
\begin{figure}[b!]
    \centering
    \includegraphics[width=0.75\textwidth]{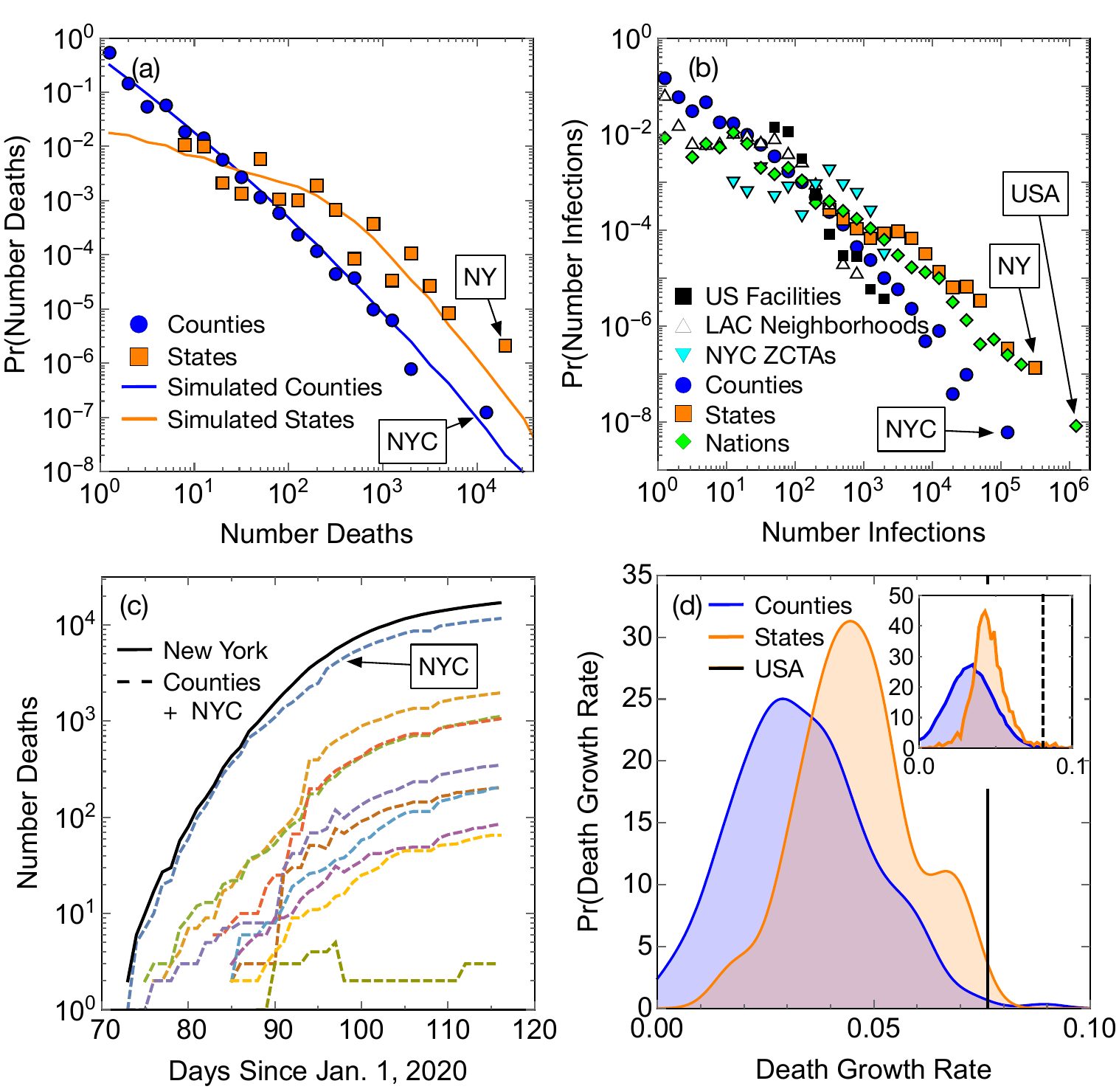}
    \caption{Inequality of COVID-19 impact and spatial aggregation bias. (a) The number of deaths has a heavy-tailed distribution for both states and counties, with the most cases in New York state and New York City, respectively. A stochastic model discussed in the main text captures the properties of the distribution. (b) Similar pattern is seen for infections at many spatial scales: from U.S. facilities to neighborhoods to nations. (c) Deaths over time for New York and some of its counties, where we see the disease arrives in counties at different times, but the initial growth rate is approximately exponential. Exponential growth is calculated between day 76 and 116 (inclusive) for New York state; the time window varies for each county. (d) The death growth rate in the U.S. is higher than most states, which is in turn higher on than growth rates in individual counties. Inset: this finding is also captured by the simulation. Findings are qualitatively similar in data of infections (see Supplementary Figure~\ref{fig:InfGrowth}). 
    }
    \label{fig:GrowthAndDist}
\end{figure}

 Figure~\ref{fig:GrowthAndDist} demonstrates the unequal impact of COVID-19 in the U.S. and the world. The number of deaths in U.S. counties and states has a heavy-tailed distribution (Fig.~\ref{fig:GrowthAndDist}a). This means that the disease's toll varies enormously between places, with many communities almost unaffected and others hit hard by the pandemic. For example, New York City accounts for the bulk of all deaths in New York state, which accounts for a large fraction of all U.S. COVID-19 deaths. The number of infections has a heavy-tailed distribution across \textit{multiple spatial scales} (Fig.~\ref{fig:GrowthAndDist}b): from large outbreaks at U.S. facilities (e.g., nursing homes, prisons and meat packing plants) catalogued by the New York Times, to Los Angeles and New York City neighborhoods, U.S. counties and states, and world nations. Despite  differences in the availability of testing, there exist strong regularities in the prevalence of outbreaks at these vastly disparate spatial scales. 

How does this large variation arise? Figure~\ref{fig:GrowthAndDist}c shows the growing toll of the disease in New York state and its ten hardest-hit counties. The growth in the number of deaths within each county (and state) in the early stages of the outbreak can be roughly modeled by an exponential, which allows us to estimate the average growth rate (agreement with exponential fits has been checked in Supplementary Figure~\ref{fig:R2Fits}). There is a fairly broad distribution of death rates for counties and states (Fig.~\ref{fig:GrowthAndDist}d). However, this by itself is not enough to explain the heterogeneity in the number of cases or deaths. We must also consider that COVID-19 appears in each subregions at different times. 

This phenomenon is closely related to the Reed-Hughes mechanism \cite{Reed2002}, which explains how 
exponentially growing populations of different ages produce a power-law distribution of population sizes. However, the Reed-Hughes mechanism specifies that populations have the same growth rate and begin growing with uniform probability in time. In contrast, the start time of the outbreak in each county is approximately normally distributed, as is the growth rate. 
To validate the modified mechanism, we create synthetic data in which simulated counties have outbreaks that start at times drawn at random from a normal distribution, with growth rates chosen from another normal distribution and coefficients of growth rates drawn from a log-normal distribution.  All distribution parameters are empirically measured from fits to data, except arrival times that are gathered directly from data. 
Synthetic outbreaks within our simulated counties follow a heavy-tailed distribution (blue line in Fig.~\ref{fig:GrowthAndDist}a) similar to the empirical distribution for counties. The situation is somewhat more complex for states. Simply dividing counties across states at random, so that each simulated state ends up aggregating data over $62$ counties (this is the mean number of counties in a state), creates a very sharply-peaked distribution, unlike what we observe. 
Instead, we divide up counties non-uniformly across states such that the number of counties in these simulated states matches the true distribution. 
This re-creates the heavy-tailed distribution of the number of deaths for states (orange line in Fig.~\ref{fig:GrowthAndDist}a). 
These results demonstrate that large heterogeneity in the data creates a qualitatively similar outcome as the Reed-Hughes mechanism.

\begin{figure}
    \centering
    \includegraphics[width=0.9\textwidth]{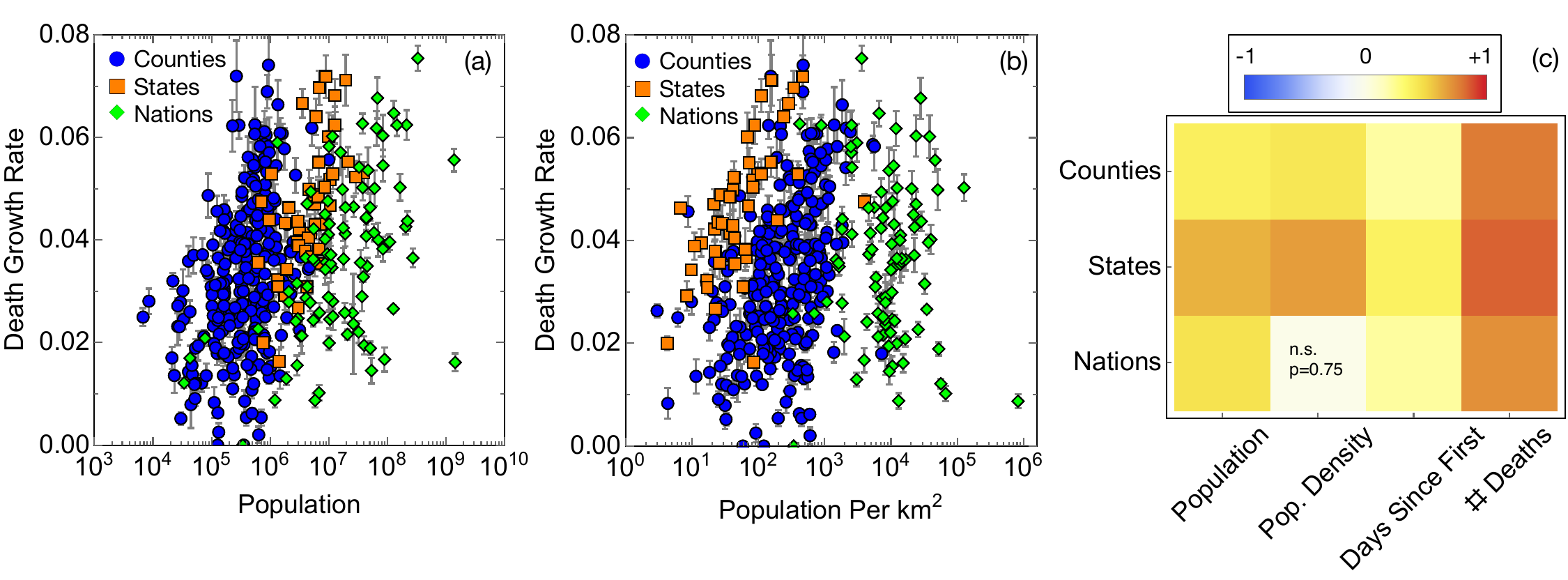}
    \caption{Correlates of COVID-19-related death rates. (a) Death rates are positively correlated with the population. (b) These correlations are often higher than correlations with population density. (c) We quantify the correlations with a heat map. Areas where deaths appeared earlier also tend to have higher growth rate. Whatever the reason for the growth rate, faster growing areas have many more cases than slower-growing areas. Similar results hold for infections as well (Fig.~\ref{fig:DeathCorrelates}).
    }
    \label{fig:DeathCorrelates}
\end{figure}

\begin{figure}[!hb]
    \centering
    \includegraphics[width=0.75\textwidth]{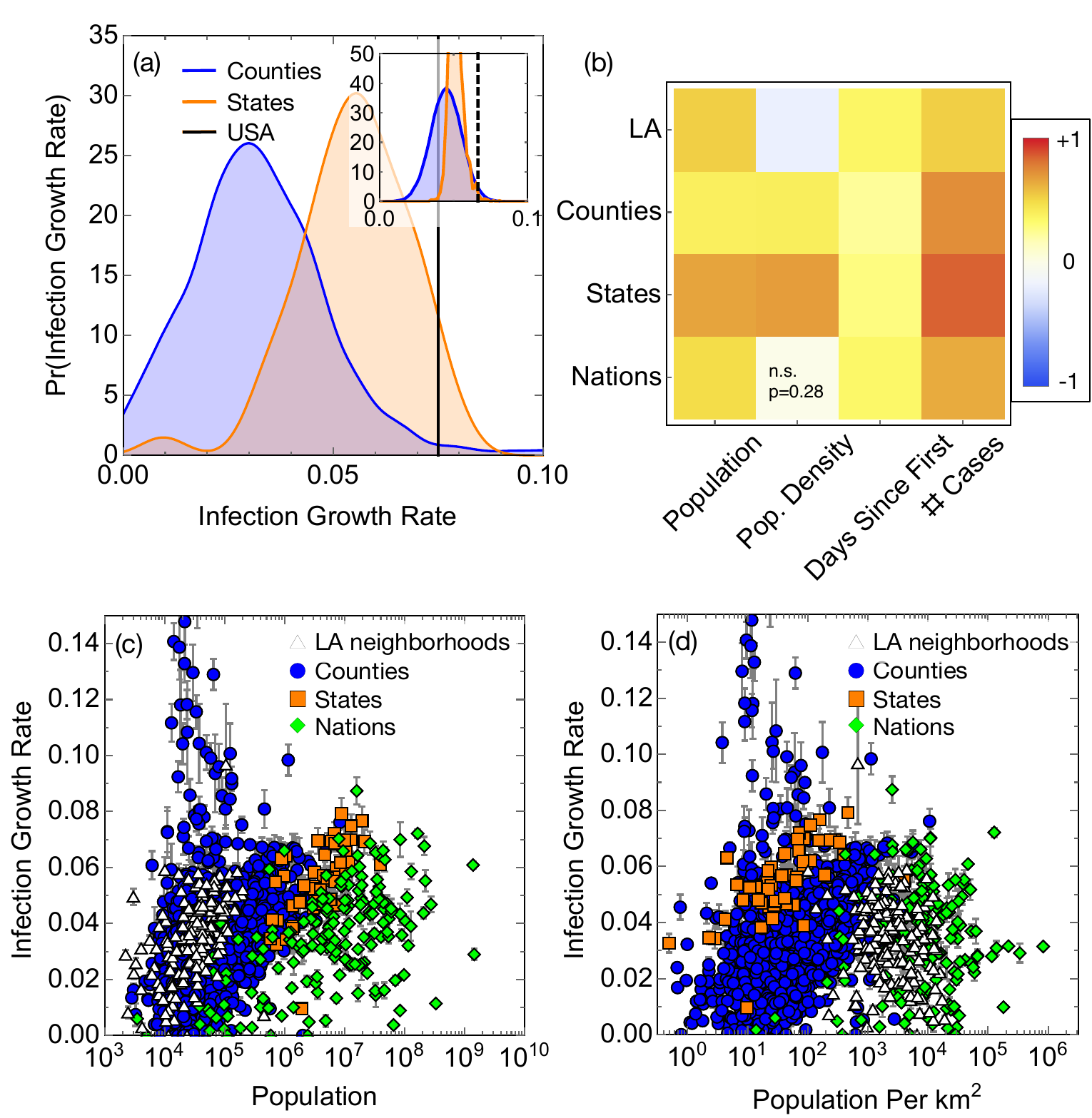}
    \caption{Infection growth. (a) We observe the infection growth rate of the U.S. is higher than most states, which is, in turn, higher than most counties. (b) We find population correlates strongly with growth rate from neighborhoods in LA to countries, but population density is less important. The time since the first infection is weakly correlated with the growth rate. Whatever the reason for the growth rate, we find the number of infections strongly correlates with the growth rate, therefore fast-growing areas dominate the statistics. Examples of growth rates versus (c) population and (d) population density.  Compare to Fig.~\ref{fig:GrowthAndDist} and~\ref{fig:DeathCorrelates} in the main text.}
    \label{fig:InfGrowth}
\end{figure}

These deviations from the traditional Reed-Hughes mechanism can also create aggregation bias. The hot spots dominate the statistics and are correlated with faster growing infections (Fig.~\ref{fig:DeathCorrelates} and~\ref{fig:InfGrowth}). This makes deaths and infections appear to grow faster when spatially aggregated, such as aggregating data from counties to the state level. Similarly, the growth rate at the national level appears to be higher than death rates within constituent states and counties (Fig.~\ref{fig:GrowthAndDist}d). Aggregation bias is also observed in simulated data with synthetic counties and states (Fig.~\ref{fig:GrowthAndDist}d inset and~\ref{fig:InfGrowth}a inset). Alike to what we observe empirically, we find Spearman correlations of 0.35 between growth rates and simulated number of deaths or growth rates and simulated numbers of infections (p-values $<10^{-6}$).

\begin{figure}[!htb]
    \centering
    \includegraphics[width=1.0\textwidth]{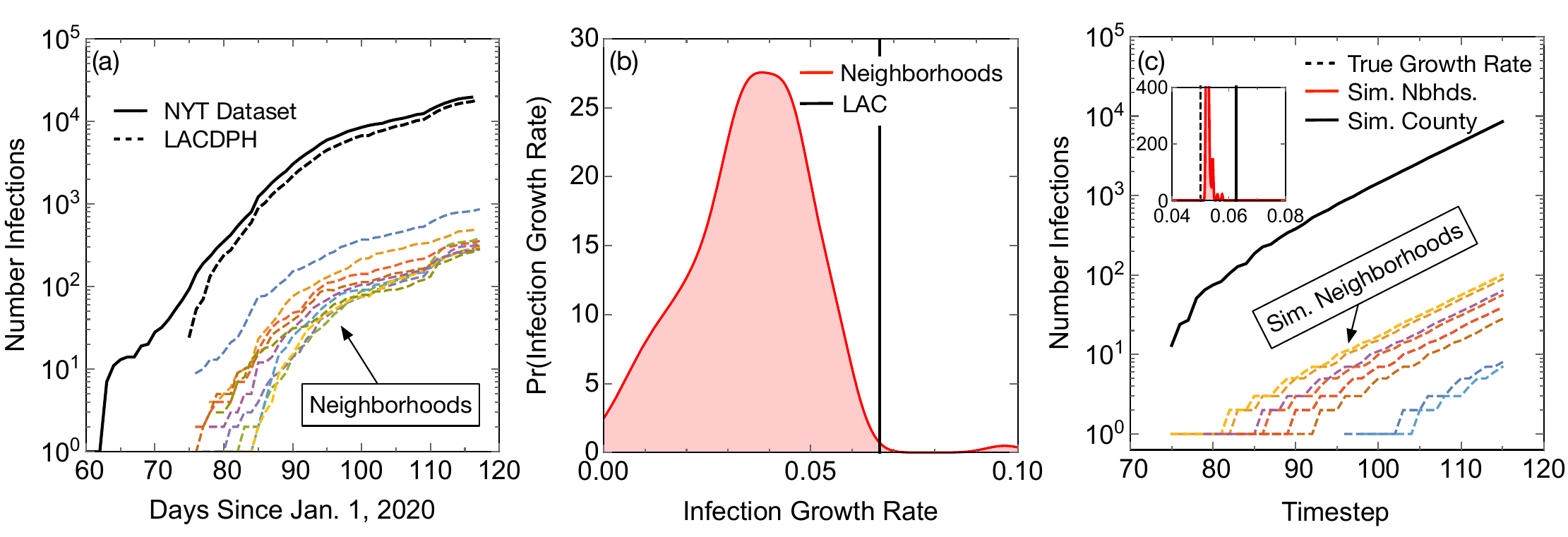}
    \caption{Growth of COVID-19 in Los Angeles. (a) We find reasonable agreement between infections based on the NYT dataset \cite{NYT2020} and the LA County Department of Public Health. Exponential growth for the county is measured between day 64 and 116 (the time window varies for each neighborhood). (b) Shows the growth rates of individual neighborhoods where the dashed line is LA County as a whole. (c) To better understand why, we make a simulation where infections grow with the same exponential growth rate ($\sim 10^{0.05 t}$ or 11.5\% a day), and the arrival time of the infection in each neighborhood matches LA data. In this idealized simulation, each neighborhood's growth rate can be easily fit to an exponential, but the aggregate fails to follow an exponential. A reasonable fit significantly over-shoots the growth rate of each neighborhood.
    }
    \label{fig:LAGrowth}
\end{figure}

Aggregation bias can also arise when combining data from communities with similar growth rates. Our analysis of comprehensive COVID-19 data from cities and neighborhoods within Los Angeles County  (Fig.~\ref{fig:LAGrowth}) shows that while the average growth rates in the number of infections within neighborhoods is similar to each other (Fig.~\ref{fig:LAGrowth}a), they are substantially lower than the growth rate for Los Angeles County as a whole (Fig.~\ref{fig:LAGrowth}b). The difference arises from the staggered arrival of the disease in different neighborhoods. To validate this, we simulated neighborhoods with the same growth rate of $\sim 10^{0.05 t}$, with arrival times taken directly from the Los Angeles neighborhoods. Even though each simulated neighborhood has the same growth rate, their staggered arrival times lead to a higher aggregate growth rate.

\begin{figure}[!htb]
    \centering
    \includegraphics[width=0.75\textwidth]{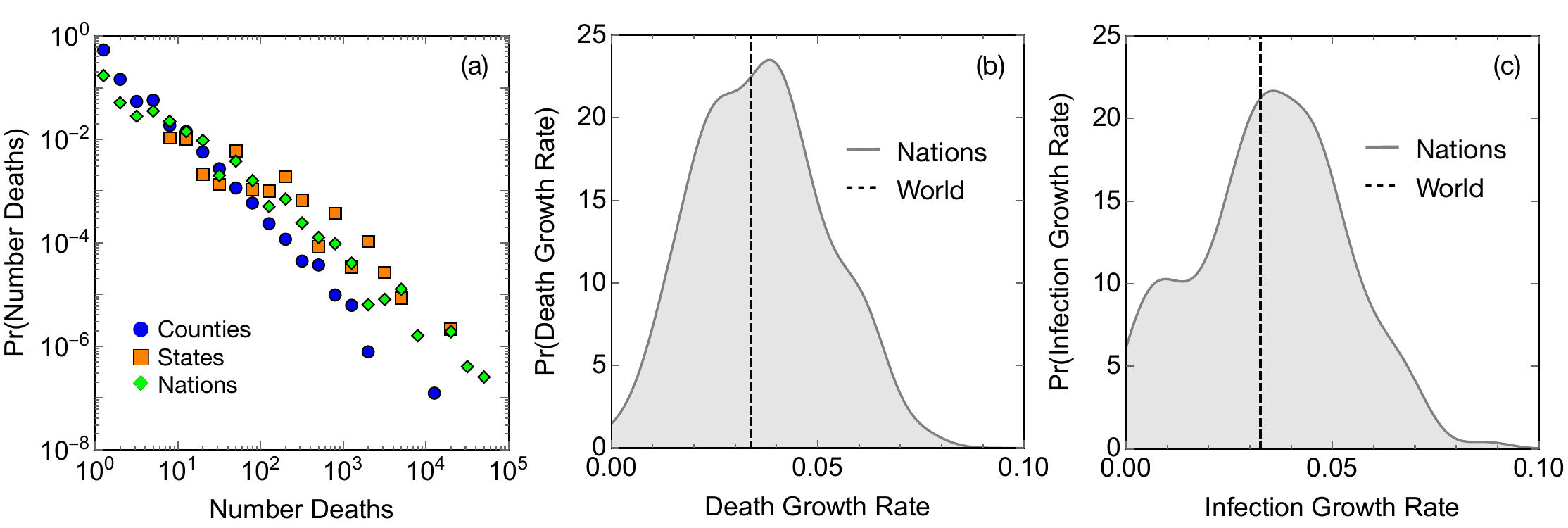}    \caption{Infections across countries. The number of infections across countries is seen in Fig.~\ref{fig:GrowthAndDist}. (a) We similarly show the distribution of the number of deaths for each country compared to counties and states in the United States. (b--c) We show the growth distribution among countries and for the world as a whole (dashed lines). Surprisingly, the world as a whole has a growth rate comparable to most countries, most likely because the growth rate is initially dominated by just one country, China, therefore the world growth is dominated by the first, rather than the fastest growing, country.
    }
    \label{fig:WorldGrowth}
\end{figure}

The systematic overestimation of aggregated growth rates is an example of Modifiable Areal Unit Problem~\cite{Gehlke1934}, a statistical bias similar to Simpson's paradox~\cite{lerman2018computational},  that results in varying statistical trends at different levels of aggregation of heterogeneous data.
Interestingly, the only scale we do not see aggregation bias is the global scale. Infections growth rate for the world as a whole is comparable to most countries (Fig.~\ref{fig:WorldGrowth}). This is, however, because the growth is initially dominated by just one country, China. The initial world growth rate is correlated to the first, rather than the fastest growing, country.



Lastly, we explore why growth rates vary. We find that growth rates are correlated with population and population density (see Figs.~\ref{fig:DeathCorrelates} and~\ref{fig:InfGrowth}): COVID-19 spreads more quickly in large and dense counties, states and nations. Population density appears to play a somewhat more important role in explaining death growth rates for states. However, this is not the case for infections, where population, rather than population density, better explains the growth of infections across multiple spatial scales (Fig.~\ref{fig:InfGrowth}c), in agreement with trends for cities reported by \cite{Stier2020}. While it seems intuitive that denser places with more interpersonal interaction are at a greater risk for spreading the disease, this may not be the entire explanation. Instead, it appears that the total number of people is an important driver. In addition, communities where the disease arrived earlier also tend to have higher growth rates, possibly because it was allowed to spread before mitigation measures were introduced. 

\section*{Discussion}


COVID-19's toll around the world varies widely, with many regions seeing few deaths and confirmed cases, while a handful of regions are greatly affected. The heavy-tailed distribution of impact has important implications for decision makers. First, local hot spots, where the virus is far more prevalent than elsewhere, are typically the more vulnerable communities with large populations. These hot spots bias aggregated growth rates COVID-19 statistics, making the disease appear to grow faster at a larger scale than it does within the constituent communities. However, we show that spatial aggregation could potentially inflate growth rates 
due to the staggered arrival of the virus in the communities even when growth rates in subregions are the same. As a result, aggregating data at a larger scale, e.g., state or national level, will make the disease appear to grow faster than it does within the constituent regions. 
%

Analysis of the effects of interventions, including lock downs and other mitigation strategies, has to account for potential biases introduced by data aggregation. Local hot spots and staggered arrival of infections may effectively amplify the rates of the disease for some regions (e.g., states and countries), obscuring the benefits of early interventions. 

From the modeling perspective, since epidemic statistics are driven by a few hot spots (typically large, dense cities or facilities),  compartmental models \cite{Coburn2009} may be more effective for modeling the disease. The assumptions made by compartmental models, namely uniform mixing of populations, are best aligned with mobility patterns in cities~\cite{Bettencourt2013} and facilities that regularly bring people in contact with one another. Compartmental models typically have fewer fitting parameters than spatio-temporal models~\cite{Brockmann2013,Burghardt2016,Liu2020}, and therefore, may be better at making intermediate-range forecasts \cite{IHME2020}. That being said, such models may produce poorer predictions due to staggered disease arrivals that spatio-temporal models can better control for.

Future work is needed to understand how these results generalize to other diseases. For example, it is important to test the Reed-Hughes-like statistical model to other diseases and countries to see the degree to which it can help explain infection hot spots. We do, however, observe some ways in which our findings differ from other diseases. For example, the growth rate of Ebola is negatively correlated with population density \cite{Burghardt2016}, potentially due to lack of healthcare infrastructure. But this may a special case, due to the impoverished countries that were infected. 

\section*{Methods and Materials}
Data on cumulative COVID-19 infections is obtained from the New York Times \cite{NYT2020} as of April 27, 2020. We also collect population and area within New York City ZCTAs, Los Angeles neighborhoods, counties, and states from the U.S. Census where population estimates are as of July, 2019 \cite{CensusArea,CensusPop}. States are defined as those with official statehood as well as the District of Columbia. Counties are defined the same as in the census except for New York City, where all boroughs are combined, and in Kansas City, where the population and area are calculated separately. Because Kansas city overlaps with other Missouri counties, we do not remove the city area from our estimates of county areas. We do not expect a significant change in our results due to this decision. 

We also collected data of U.S. facilities as of May 5th from New York Times \cite{NYTclusters}, and infections from New York ZCTAs as of April 25th from the NYC department of health \cite{NYCH2020}, and from Los Angeles County neighborhoods as of April 27th from the Los Angeles County Department of Public Health \cite{LADPH}. Finally, we collected data across nations from Our World In Data as of April 29th \cite{OWD2020}. Populations and areas of each nation were gathered from the United Nations \cite{UNDataPop,UNDataLand}.

Growth rates are calculated by taking the log base 10 of the cumulative infections (and deaths) and fitting a line. For these all fits except Los Angeles neighborhoods, data below 11 infections (deaths) are removed to reduce effects of outliers. The threshold is three  for Los Angeles neighborhood data because agreement with exponential is reasonable even when we start with this lower threshold. Over time the data deviates from exponential growth. We consider the data to deviate significantly if we have more than two consecutive days of growth below 0.5\%. Those two days, and all subsequent data, is removed from analysis. A more stringent threshold of 0.1\% growth over three days produces very similar results. In addition, we only fit data with more than five datapoints. Calculations of $R^2$ are based on this log-scaling of data.  

\section*{Supplementary Figures}

\begin{figure}[!htb]
    \centering
    \includegraphics[width=0.75\textwidth]{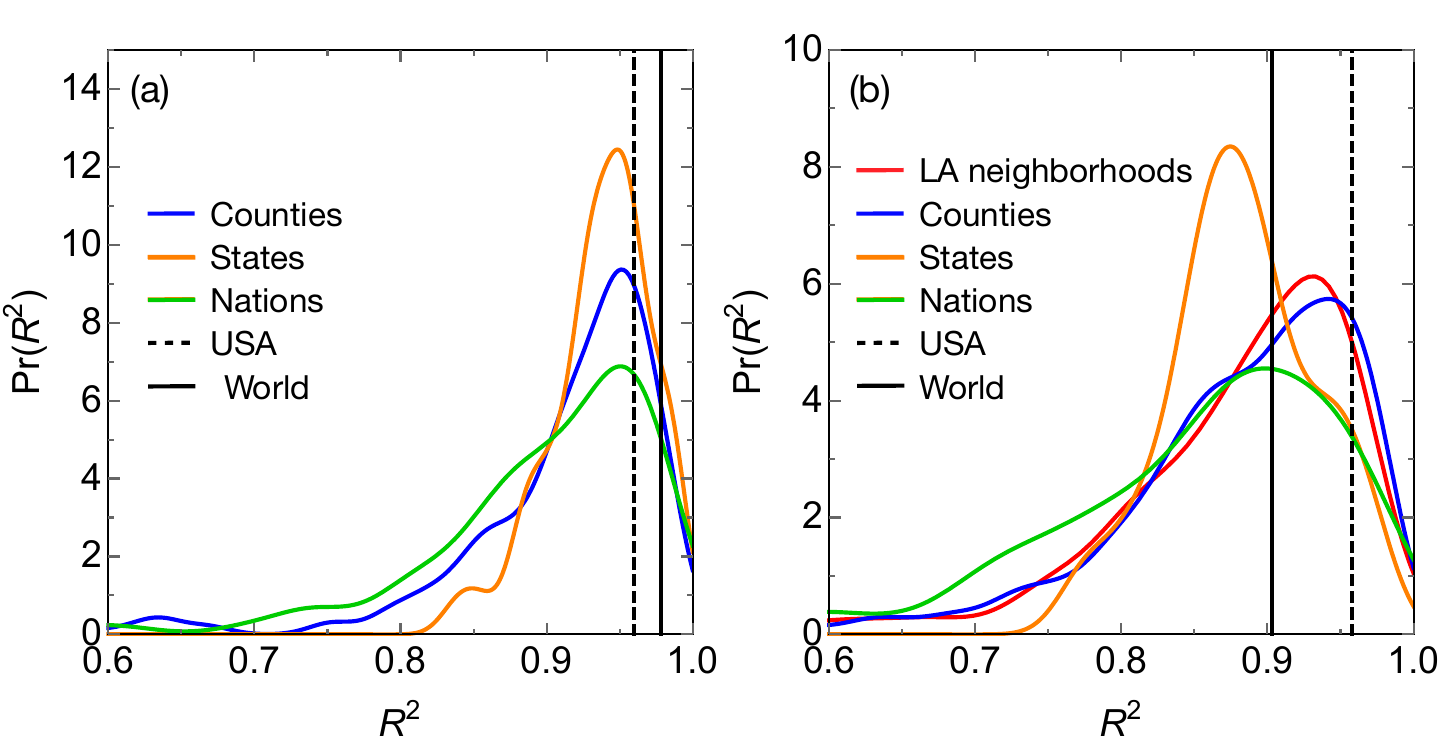}    \caption{Quality of the exponential fits. We measure fit quality using $R^2$ for a linear fit of log-scaled data. For (a) deaths and (b) infections, we see a large majority of fits have extremely high $R^2$.}
    \label{fig:R2Fits}
\end{figure}

\section*{Acknowledgments}
This work was funded in part by DARPA under contract HR00111990114.

\end{document}